\documentclass[12pt]{iopart}
\usepackage{iopams} 
\usepackage{graphicx} 
\usepackage{caption}
\usepackage{wrapfig}
\usepackage{booktabs}
\usepackage{cleveref}
\usepackage{url}
\usepackage{xcolor}
\usepackage{multirow}

\begin{document}

\title[Manuscript]{Validation of neutron emission and neutron energy spectrum calculations on MAST with DRESS}

\author{A. Sperduti$^1$, I. Klimek$^1$, S. Conroy$^1$, M. Cecconello$^1$,\\
	  M. Gorelenkova$^2$ and A. Snicker$^3$}

\address{$^1$ Department of Physics and Astronomy, Uppsala University, SE-751 05 Uppsala, Sweden}
\address{$^2$ Princeton Plasma Physics Laboratory, Princeton, NJ 08543-0451, USA}
\address{$^3$ Department of Applied Physics, Aalto University, P.O. Box	11100, 00076 AALTO, Finland} 
 
\ead{andrea.sperduti@physics.uu.se}

\begin{abstract}
The recently developed Directional RElativistic Spectrum Simulator (DRESS) code has been validated for the first time against numerical calculations and experimental measurements performed on MAST. In this validation, the neutron emissivities and rates computed by DRESS are benchmarked against TRANSP/NUBEAM predictions while the neutron energy spectra provided by DRESS taking as input TRANSP/NUBEAM and ASCOT/BBNBI in Gyro-Orbit (GO) mode fast ion distributions are validated against proton pulse height spectra (PHS) measured by the neutron flux monitor. Excellent agreement was found between DRESS and TRANSP/NUBEAM predictions of local and total neutron emission.
\end{abstract}

\section{Introduction}
Modelling of the neutron emission from the plasma can be used to assess the local and total plasma performances in terms of fast ions confinement and transport while modelling of the neutron energy spectrum can give insight into the velocity distributions of the interacting fuel ions. Plasma modelling codes such as TRANSP~\cite{TRANSP} and injected neutral beam deposition and slowing down codes such as NUBEAM~\cite{NUBEAM} are typically used to model the neutron emissivity and rate which are then compared with and validated against global measurements made usually with calibrated fission chambers and multi-chord neutron flux monitors. However, neither TRANSP nor NUBEAM calculate the neutron energy spectrum in a specific direction of observation and field of view of a neutron diagnostic. A Directional RElativistic Spectrum Simulator (DRESS) code~\cite{DRESS} has been developed to model the energy spectra of the products from fusion reactions involving two reactants with arbitrary velocity distributions for arbitrary observational directions. DRESS employs fully relativistic kinematic equations to calculate the energy of the reactions' products. The only inputs to DRESS are the type of reacting fuel ions, their velocity distributions and the equilibrium magnetic field. The observational direction is provided by the specific neutron diagnostic for which the modelled neutron emission and spectrum are requested.\\
So far, DRESS has been successfully benchmarked against analytical approximations~\cite{DRESS} and it has been used to generate plasma neutron sources for neutron transport calculations on JET \cite{ZIGA1, ZIGA2}. In this work, the first validation of DRESS against numerical and experimental measurements on the Mega Ampere Spherical Tokamak (MAST)~\cite{MAST} is presented. TRANSP and NUBEAM are commonly used at MAST to model the global neutron rate which is compared with the one measured by an absolutely calibrated fission chamber (FC)~\cite{FC}. In addition, the TRANSP/NUBEAM predicted neutron emissivities are used for the forward modelling of neutron count profiles measured by a neutron profile monitor commonly called Neutron Camera (NC)~\cite{NC}. The NC has two equatorial and two diagonal collimated Lines of Sight (LoS) equipped with EJ301 liquid scintillation detectors. The NC measures, in a single plasma discharge, the neutron count rate at four positions characterized by two tangency radii $p_1$ and $p_2$ and two vertical coordinates $Z_1 =0$ and $Z_2 = -20$ cm. Profiles are obtained by moving the NC in between repeated plasma discharges. \\
In MAST, the agreement between the forward modelled NC count rates using TRANSP/NUBEAM and NC measurements requires the multiplication of the predicted rate by a scaling factor of $k$ = 0.6 $\pm$ 0.1, which is independent of the plasma scenario. The origin of this discrepancy has been investigated in details in~\cite{MASTDEFICIT}. In that work it was concluded that experimental uncertainties in the plasma parameters in input to TRANSP/NUBEAM that affect the neutron emission could not account for the 40 \% discrepancy. Detailed analysis of further possible sources for this discrepancy will be carried out in an accompanying paper: among them the Guiding Center (GC) approximation used by TRANSP/NUBEAM is tested in this work by calculating the neutron energy spectra using DRESS and taking as input fast ion distributions from ASCOT/BBNBI \cite{ASCOT4, BBNBI} in Gyro Orbit (GO) and from TRANSP/NUBEAM in GC.\\
This study is based on one of the scenarios studied in~\cite{MASTDEFICIT, Klimek}, which is characterized by a plasma current of $780$ kA, a neutral beam injection (NBI) heating of $1.6$ MW and a total neutron rate of ${\mathrm{Y_n}} \simeq 3\times 10^{13}\ \mathrm{s}^{-1}$. This is an MHD-quiescent plasma, free of anomalous fast ion losses that might affect neutron emission, hence ideal for benchmarking of neutron emissivities and rates calculated by DRESS against TRANSP/NUBEAM predictions.\\
This paper is structured as follows. Neutron emissivities and rates calculated by DRESS starting from TRANSP/NUBEAM fast ion distribution are shown and discussed in section \ref{sec:NeutronEmissivityRate}. The neutron energy spectrum at the NC detector location computed by DRESS starting from TRANSP/NUBEAM and ASCOT/BBNBI fast ion distributions are convoluted with the NC detectors' response matrix and the resulting recoil proton pulse height spectra (PHS) are then compared with the measured ones as described in section \ref{sec:NeutronSpectra} of this paper. Finally, the summary and the conclusions are described in section \ref{conclusions}.

\section{Neutron emissivities and rates}
\label{sec:NeutronEmissivityRate}
The number of neutrons emitted per unit volume and time (also called the emissivity $\xi$) from a plasma containing ion species of type ``a'' and ``b'' is calculated according to
\begin{equation}
\xi_{\mathrm{ab}}(\mathrm{R,Z};\rho_{\phi}) =  \frac{n_{\mathrm{a}}n_{\mathrm{b}}}{1 + \delta_{\mathrm{a}\mathrm{b}}} \langle \sigma v \rangle_{\mathrm{ab}}     
\label{emissivity}
\end{equation}
where $\rho_{\phi}$ denotes the normalized toroidal flux, R and Z the coordinates of a point in the poloidal cross-section, $n_{\mathrm{a}}$ and $n_{\mathrm{b}}$ are the fuel ion densities of species ``a'' and ``b'', $\delta_{\mathrm{a}\mathrm{b}}$ is the Kronecker's delta included in order not to double count ions belonging to the same population and $\langle \sigma v \rangle$ is the reactivity between the two species. Since MAST operated only in plasmas with D fuel ions and D neutral beam, $\sigma$ indicates unambiguously the DD cross-section.\\
In a thermal plasma, fuel ions have a Maxwell-Boltzmann velocity distribution and an analytical expression for the thermonuclear reactivity $\left\langle  \sigma v \right\rangle_{\mathrm{th}}$ can be derived. The most accurate parametrization of $\left\langle  \sigma v \right\rangle_{\mathrm{th}}$ can be found in equation 12 of \cite{BOSCH_HALE}. Similarly, semi-analytical approximations for $\langle\sigma v\rangle$ may be derived for plasmas in which ions have non-thermal velocity distributions originating, for example, from ion cyclotron resonance or neutral beam heating \cite{GONCHA}. These approximations are typically not accurate enough for most applications and therefore full numerical calculations involving integration over velocity distributions are performed in codes such as NUBEAM, DRESS, FPS \cite{FPS}, FPSLOS \cite{FPSLOS}, LINE \cite{LINE}, GENESIS \cite{GENESIS}, ControlRoom \cite{CONTROLROOM} and AFSI \cite{AFSI}. It is worth mentioning that the aforementioned codes can also evaluate spectra for a LoS or a point in the plasma. The main advantage of DRESS respect to the other codes is the implementation of relativistic calculations whilst codes such as AFSI, GENESIS and ControlRoom are non-relativistic and operate in centre-of-mass coordinates.

\subsection{Neutron emissivity calculations in TRANSP/NUBEAM}  

\begin{figure}[!bth]
\centering
\includegraphics[scale=0.6, trim = 0cm 0cm 0cm 0cm,]{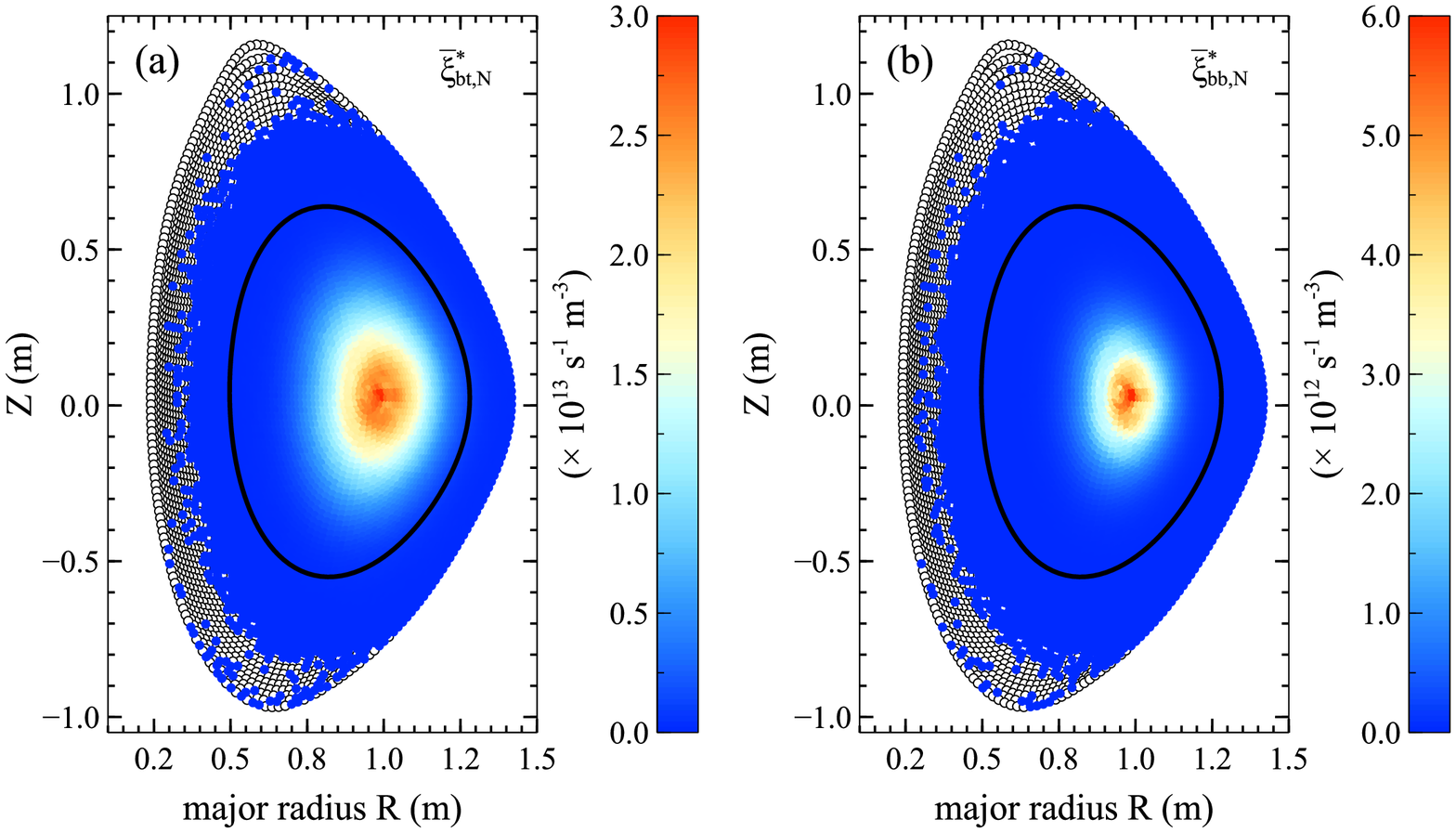}
\includegraphics[scale=0.6, trim = 0cm 0cm 0cm 0cm,]{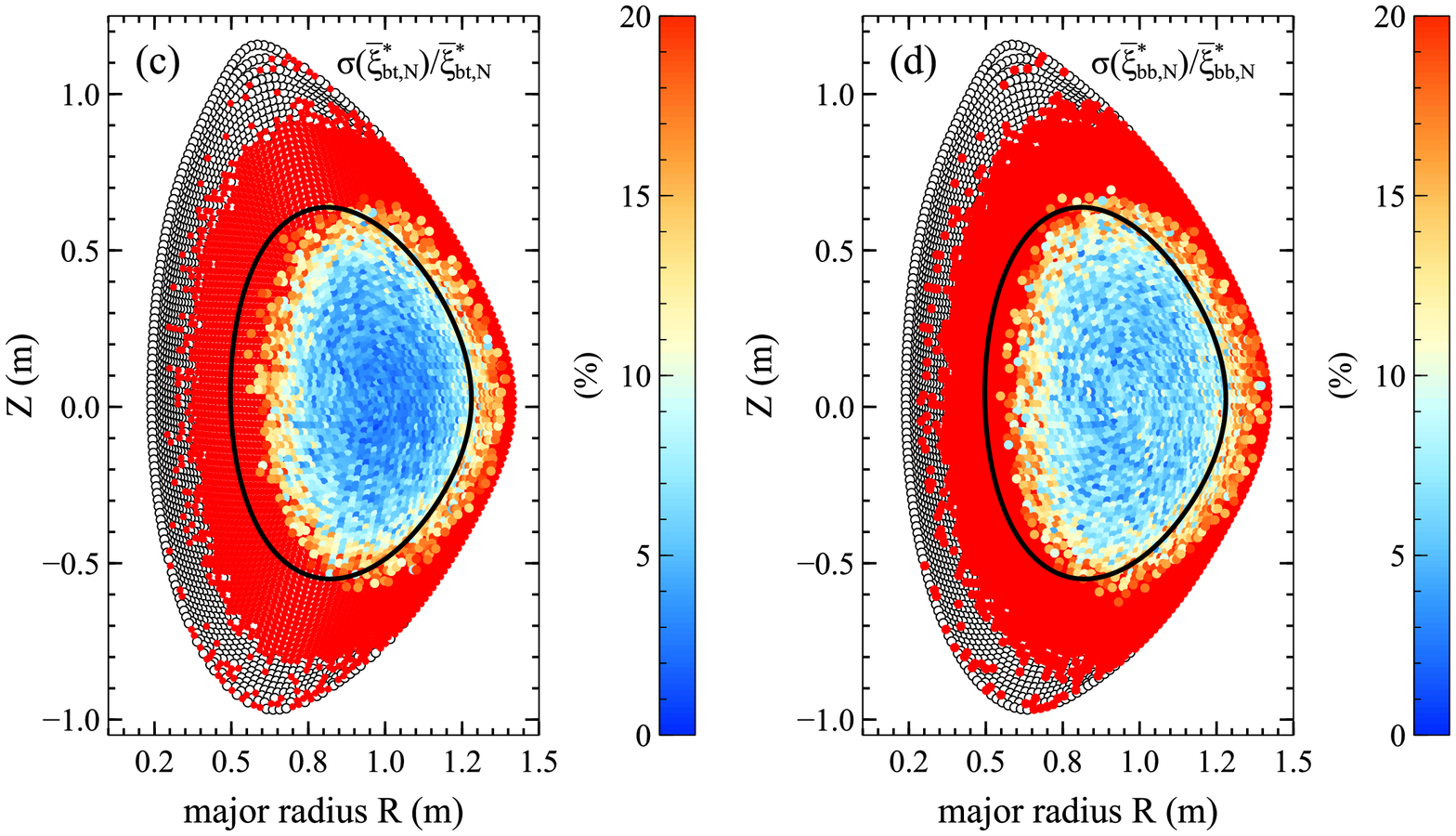}
\caption{Pulse 29909: a) $\overline{\xi}^*_{\mathrm{bt,N}}$, b) $\overline{\xi}^*_{\mathrm{bb,N}}$, c) $\sigma(\overline{\xi}^*_{\mathrm{bt,N}}) / \overline{\xi}^*_{\mathrm{bt,N}}$ and d) $\sigma(\overline{\xi}^*_{\mathrm{bb,N}}) / \overline{\xi}^*_{\mathrm{bb,N}}$ at $t = 0.216$ s. The flux surface with $\rho_{\phi} = 0.575$ shown in black encloses the plasma region from which $99\%$ of all neutrons are emitted. The fractional errors exceeding $20\%$ are plotted in red. Spatial points from which no neutrons are emitted are plotted in white.}\label{fig:2D_BT_BB_TR_mean_realtive_error}
\end{figure}

TRANSP and NUBEAM codes use equation~\ref{emissivity} to compute neutron emissivities on different but spatially aligned grids. In TRANSP, the plasma is divided into a number of annular regions constrained by surfaces equally spaced in normalized toroidal flux $\rho_{\phi}$ creating a 1D grid. Thermal ion density $n_{\mathrm{d}}$ and temperature $T_{\mathrm{i}}$ are considered to be flux-surface quantities in TRANSP, hence the thermonuclear emissivity $\xi_{\mathrm{th,T}}$ computed by TRANSP (index ``$\mathrm{T}$'' denotes TRANSP) using directly the parametrization of $\left\langle  \sigma v \right\rangle _{\mathrm{th}}$ from~\cite{BOSCH_HALE} is also a flux-surface quantity. It is worth noting that TRANSP computes $\xi_{\mathrm{th,T}}$ for $T_{\mathrm{i}}$ in the range of $0.2$ to $100$ keV (in which the parametrization of $\left\langle  \sigma v \right\rangle _{\mathrm{th}}$ from~\cite{BOSCH_HALE} is valid) and sets $\xi_{\mathrm{th,T}}$ to zero for $T_{\mathrm{i}}$ outside this range. In NUBEAM, the annular regions are additionally subdivided into a number of poloidal zones, whose number increases with $\rho_{\phi}$ creating a 2D spatial grid. The 4D fast ion velocity distribution function at time $t$ for each poloidal zone individuated by the zone's centroid coordinates $(\mathrm{R,Z})$,  $f(\mathrm{R,Z},E, \lambda)$, where $E$ is the fast ion energy and $\lambda$ is the ratio between the fast ion velocity $v$ and its parallel component $v_{||}$, is used to calculate the local beam-thermal $\xi^*_{\mathrm{bt,N}}$ and beam-beam $\xi_{\mathrm{bb,N}}^*$ neutron emissivities (where index ``N'' denotes NUBEAM and ``$^*$'' refers to the non-flux averaged neutron emissivity).\\
The flux-averaged beam-thermal $\xi_{\mathrm{bt,T}}$ and beam-beam $\xi_{\mathrm{bb,T}}$ neutron emissivities, obtained by averaging $\xi^*_{\mathrm{bt,N}}$ and $\xi^*_{\mathrm{bb,N}}$ calculated by NUBEAM over poloidal zones in each flux surfaces and by mapping them onto the 1D TRANSP grid by NUBEAM, are then added to $\xi_{\mathrm{th,T}}$ to obtain the total neutron emissivity. Similarly, the beam-thermal $R_{\mathrm{bt,T}}$ and the beam-beam $R_{\mathrm{bb,T}}$ neutron rates, obtained by integration of $\xi^*_{\mathrm{bt,N}}$ and $\xi^*_{\mathrm{bb,N}}$ over the plasma volume, are then added to $R_{\mathrm{th,T}}$ in order to calculate the total neutron rate $R_{\mathrm{T}}$. The quantities $\xi_{x,\mathrm{T}}$ and $R_{x,\mathrm{T}}$ are calculated for all simulation time steps (here $x$ stands for ``th'', ``bt'' or ``bb''). On the other hand, $f(\mathrm{R,Z},E, \lambda)$ along with $\xi^*_{\mathrm{bt,N}}$ and $\xi^*_{\mathrm{bb,N}}$ calculated by NUBEAM at each simulation time step cannot be directly accessed. However, $\xi^*_{\mathrm{bt,N}}$ and $\xi^*_{\mathrm{bb,N}}$ are necessary for the proper modelling of NC count rate profiles as reported in~\cite{Klimek}. NUBEAM was therefore modified to store, in an output file accessible to the user, the time-averaged local $\langle\xi^*_{\mathrm{bt,N}}\rangle$ and $\langle\xi^*_{\mathrm{bb,N}}\rangle$, where '$\langle \rangle$' denotes the average over a given time interval $\Delta t$ specified by the user. The calculated $\langle\xi^*_{\mathrm{bt,N}}\rangle$ and $\langle\xi^*_{\mathrm{bb,N}}\rangle$ are then added to $\langle\xi_{\mathrm{th,T}}\rangle$ to obtain the total non-flux averaged neutron emissivity. In this work, $\langle\xi^*_{\mathrm{x,N}}\rangle \equiv  \xi^*_{\mathrm{x,N}}$ and $\langle\xi_{\mathrm{th,T}}\rangle \equiv \xi_{\mathrm{th,T}}$ as both the time interval $\Delta t$ and the simulation time $t_{\mathrm{k}}$ step were set equal to 1 ms.\\
Since NUBEAM is a Monte Carlo code, the computed $f(\mathrm{R,Z},E, \lambda)$, emissivities and rates are subject to statistical fluctuations. However NUBEAM does not provide an estimate of the statistical uncertainties for these quantities. In order to estimate them, seven TRANSP/NUBEAM simulations have been performed using the same input data to obtain estimates of the mean beam-thermal $\overline{\xi}^*_{\mathrm{bt,{N}}}$ and beam-beam $\overline{\xi}^*_{\mathrm{bb,{N}}}$ emissivities along with the corresponding standard deviations $\sigma(\overline{\xi}^*_{\mathrm{bt,{N}}})$ and $\sigma(\overline{\xi}^*_{\mathrm{bt,{N}}})$ (where apex ``$^-$'' denotes the average over the seven repeated TRANSP/NUBEAM simulations). The calculated $\overline{\xi}^*_{\mathrm{bt,{N}}}$ and $\overline{\xi}^*_{\mathrm{bb,{N}}}$ along with the estimated fractional errors are presented in figure \ref{fig:2D_BT_BB_TR_mean_realtive_error}. Almost all beam-thermal and beam-beam neutrons $(\simeq 99\%)$ are emitted from the plasma region enclosed by the flux surface characterized by $\rho_{\phi} = 0.575$ and shown in black in figure \ref{fig:2D_BT_BB_TR_mean_realtive_error}. The fractional errors exceeding $20\%$ are plotted in red while spatial points from which no neutrons are emitted are shown in white. As can be seen, the fractional errors are generally smaller than $10\%$ within $\rho_{\phi} \leq 0.575$, although there is a narrow annular region for $\rho_{\phi} \simeq 0.575$ where the fractional errors exceed $20\%$. For $\rho_{\phi} >0.575$, the fractional errors on $\overline{\xi}^*_{\mathrm{bt,{N}}}$ and $\overline{\xi}^*_{\mathrm{bb,{N}}}$ are large due to large statistical fluctuations in the computed $f(\mathrm{R,Z}, E, \lambda)$.  

\begin{figure}[!bth]
\centering
\includegraphics[scale=0.75]{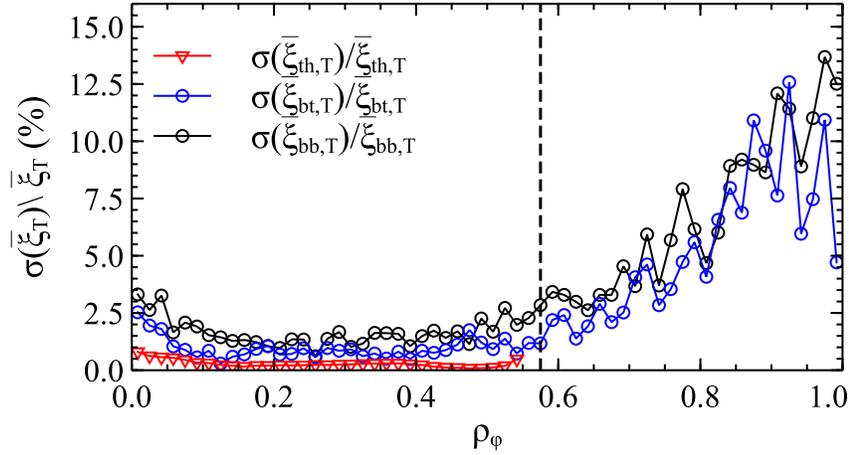} 
\caption{Pulse 29909: fractional error on the $\overline{\xi}_{\mathrm{bt,T}}$, $\overline{\xi}_{\mathrm{bb,T}}$ and $\overline{\xi}_{\mathrm{th,T}}$ neutron emissivity components estimated from the seven TRANSP simulations. The dashed vertical line marks $\rho_{\phi} = 0.575$.}\label{fig:Ratio_FAEmissivity}
\end{figure}

In addition, the mean flux-averaged beam-thermal $\overline{\xi}_{\mathrm{bt,T}}$, beam-beam $\overline{\xi}_{\mathrm{bb,T}}$ and thermonuclear $\overline{\xi}_{\mathrm{th,T}}$ emissivities together with the corresponding standard deviations $\sigma(\overline{\xi}_{\mathrm{bt,T}})$, $\sigma(\overline{\xi}_{\mathrm{bb,T}})$ and $\sigma(\overline{\xi}_{\mathrm{th,T}})$ have been calculated. The fractional errors on the  $\overline{\xi}_{\mathrm{bt,T}}$, $\overline{\xi}_{\mathrm{bb,T}}$ and $\overline{\xi}_{\mathrm{th,T}}$ are shown in figure~\ref{fig:Ratio_FAEmissivity}. The smallest fractional error is observed for $\overline{\xi}_{\mathrm{th,T}}$. In the plasma region up to $\rho_{\phi} \leq 0.575$, the small statistical fluctuations on the $\xi^*_{\mathrm{bt,N}}$ and $\xi^*_{\mathrm{bb,N}}$ are additionally smoothed out by averaging over poloidal zones in each flux surface, resulting in a fractional error on $\overline{\xi}_{\mathrm{bt,T}}$ and $\overline{\xi}_{\mathrm{bb,T}}$ less than $3\%$ and $4\%$, respectively. For $\rho_{\phi} \geq 0.575$, the fractional errors on the $\overline{\xi}_{\mathrm{bt,T}}$ and $\overline{\xi}_{\mathrm{bb,T}}$ increase as a result of much larger statistical fluctuations on $\xi^*_{\mathrm{bt,N}}$ and $\xi^*_{\mathrm{bb,N}}$. 

\begin{figure}[!bth]
\centering
\includegraphics[scale=0.75]{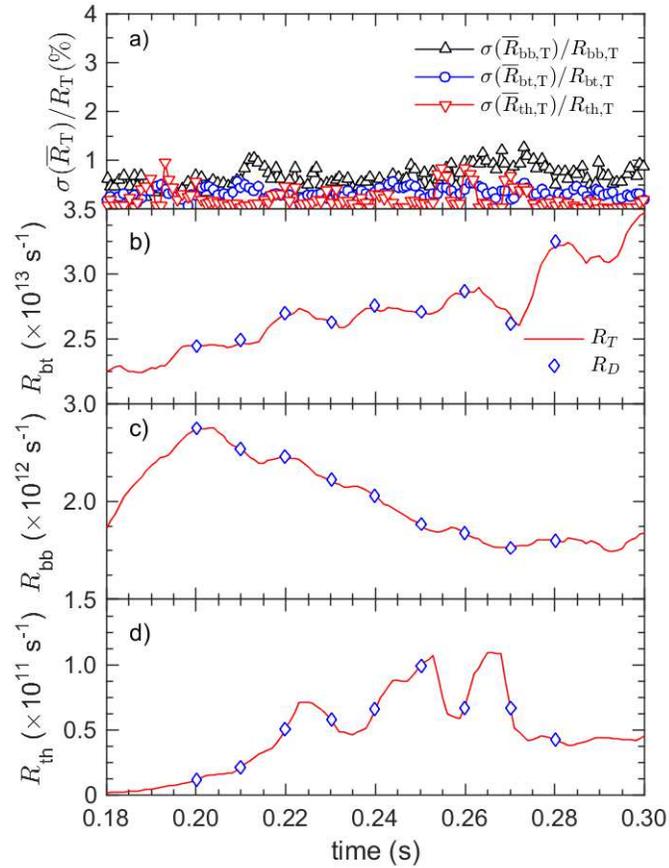}
\caption{Pulse 29909: a) the fractional errors on $R_{\mathrm{bt,T}}$, $R_{\mathrm{bt,T}}$ and $R_{\mathrm{th,T}}$ estimated from the seven identical TRANSP simulations, b) $R_{\mathrm{bt,T}}$, c) $R_{\mathrm{bb,T}}$ and d) $R_{\mathrm{th,T}}$ for one TRANSP/NUBEAM simulation. The $R_{\mathrm{bt,D}}$, $R_{\mathrm{bb,D}}$ and $R_{\mathrm{th,D}}$ for selected times are plotted as blue diamonds. The red lines represent the neutron rates $R_T$ calculated by TRANSP/NUBEAM.}\label{fig:Rates}
\end{figure}

Finally, the fractional errors on the thermonuclear $R_{\mathrm{th,T}}$, beam-thermal $R_{\mathrm{bt,T}}$ and beam-beam $R_{\mathrm{bb,T}}$ neutron rates, estimated using the data from the seven identical TRANSP simulations, have been calculated. The absolute values of the $R_{\mathrm{th,T}}$, $R_{\mathrm{bt,T}}$ and $R_{\mathrm{bt,T}}$ obtained from one TRANSP/NUBEAM simulation for several selected time slices are presented in table~\ref{tab:Rates}. The statistical fluctuations on the predicted rates are lower than $2\%$ as can be seen in figure~\ref{fig:Rates} and table~\ref{tab:Rates}. \\
If TRANSP/NUBEAM is able to calculate internally the non-flux and the flux averaged neutron emissivities, ASCOT needs to be coupled to AFSI \cite{AFSI}. ASCOT, for a given magnetic equilibrium and kinetic profiles, is capable by means of BBNBI to calculate the injected neutrals and the fast ion birth locations  \cite{BBNBI}. ASCOT then computes the fast ion distribution following the ionized fast ions both in GC and GO until they slow down to energies below 3$T_\mathrm{i}$/2 (where $T_\mathrm{i}$ is the plasma bulk temperature) or collide with the first wall. The fast ion distribution is calculated on a uniform rectangular grid and then is passed to DRESS which calculates the components of the neutron emissivity components and the neutron spectra for a given detector location. 

\begin{table}[!bth]
	\caption{Neutron rates predicted by TRANSP and DRESS along with the estimated uncertainties.}
\small
\begin{center}
\begin{tabular}{ccccccccccccr}
\toprule
\small

time  & $R_{\mathrm{th,T}}$ & $R_{\mathrm{bt,T}}$ & $R_{\mathrm{bb,T}}$ & $R_{\mathrm{T}}$ \\
(s)  & $(\times10^{11}\ \mathrm{s^{-1}})$ & $(\times10^{13}\ \mathrm{s^{-1}})$  & $(\times10^{12}\ \mathrm{s^{-1}})$  & $(\times10^{13}\ \mathrm{s^{-1}})$ \\

\midrule

0.20 & $1.152\pm0.007$ & $2.442\pm0.007$ & $2.744\pm0.018$ &  $2.717\pm0.009$ \\
0.21 & $2.176\pm0.001$ & $2.469\pm0.011$ & $2.535\pm0.015$ &  $2.724\pm0.012$ \\
0.22 & $5.112\pm0.002$ & $2.683\pm0.008$ & $2.449\pm0.007$ &  $2.933\pm0.008$ \\
0.23 & $5.780\pm0.001$ & $2.645\pm0.005$ & $2.224\pm0.012$ &  $2.873\pm0.005$ \\
0.24 & $6.615\pm0.002$ & $2.725\pm0.010$ & $2.056\pm0.007$ &  $2.938\pm0.010$ \\
0.25 & $9.902\pm0.001$ & $2.716\pm0.008$ & $1.775\pm0.009$ &  $2.903\pm0.008$ \\
0.26 & $6.729\pm0.006$ & $2.842\pm0.012$ & $1.688\pm0.016$ &  $3.017\pm0.013$ \\
0.27 & $6.689\pm0.005$ & $2.681\pm0.005$ & $1.535\pm0.012$ &  $2.841\pm0.006$ \\
0.28 & $4.306\pm0.001$ & $3.207\pm0.005$ & $1.607\pm0.008$ &  $3.372\pm0.005$ \\
\midrule
 & $R_{\mathrm{th,D}}$ & $R_{\mathrm{bt,D}}$ & $R_{\mathrm{bb,D}}$ & $R_{\mathrm{D}}$ \\
 & $(\times10^{11}\ \mathrm{s^{-1}})$ & $(\times10^{13}\ \mathrm{s^{-1}})$  & $(\times10^{12}\ \mathrm{s^{-1}})$  & $(\times10^{13}\ \mathrm{s^{-1}})$ \\
\midrule
0.20 & $1.1500\pm0.0001$ & $2.44429\pm0.00013$ & $2.74398\pm0.00022$ &  $2.71984\pm0.00013$ \\
0.21 & $2.1703\pm0.0011$ & $2.48605\pm0.00014$ & $2.53734\pm0.00020$ &  $2.74196\pm0.00014$ \\
0.22 & $5.1100\pm0.0023$ & $2.69246\pm0.00015$ & $2.45214\pm0.00019$ &  $2.94279\pm0.00015$ \\
0.23 & $5.7942\pm0.0025$ & $2.62891\pm0.00015$ & $2.22572\pm0.00017$ &  $2.85728\pm0.00015$ \\
0.24 & $6.6003\pm0.0029$ & $2.75036\pm0.00015$ & $2.06028\pm0.00016$ &  $2.96298\pm0.00015$ \\
0.25 & $9.9096\pm0.0041$ & $2.71474\pm0.00015$ & $1.76741\pm0.00014$ &  $2.90139\pm0.00016$ \\
0.26 & $6.7277\pm0.0031$ & $2.87102\pm0.00016$ & $1.67707\pm0.00013$ &  $3.04545\pm0.00016$ \\
0.27 & $6.7098\pm0.0031$ & $2.61282\pm0.00015$ & $1.53244\pm0.00012$ &  $2.77278\pm0.00015$ \\
0.28 & $4.3038\pm0.0022$ & $3.25065\pm0.00018$ & $1.60640\pm0.00012$ &  $3.41560\pm0.00018$ \\
\midrule
 & $R_\mathrm{th,T}/R_\mathrm{th,D}$ & $R_\mathrm{bt,T}/R_\mathrm{bt,D}$ & $R_\mathrm{bb,T}/R_\mathrm{bb,D}$ & $R_\mathrm{T}/R_\mathrm{D}$ \\
\midrule
0.20 & $1.0017\pm0.0057$ & $0.9989\pm0.0028$ & $0.9999\pm0.0067$ &  $0.9990\pm0.0031$ \\
0.21 & $1.0028\pm0.0001$ & $0.9930\pm0.0044$ & $0.9991\pm0.0058$ &  $0.9935\pm0.0043$ \\
0.22 & $1.0004\pm0.0047$ & $0.9965\pm0.0028$ & $0.9989\pm0.0028$ &  $0.9967\pm0.006$ \\
0.23 & $0.9976\pm0.0014$ & $1.0061\pm0.0020$ & $0.9991\pm0.0055$ &  $1.0056\pm0.0018$ \\
0.24 & $1.0022\pm0.0031$ & $0.9909\pm0.0035$ & $0.9979\pm0.0031$ &  $0.9914\pm0.0034$ \\
0.25 & $0.9993\pm0.0013$ & $1.0003\pm0.0029$ & $1.0042\pm0.0050$ &  $1.0006\pm0.0029$ \\
0.26 & $1.0002\pm0.0084$ & $0.9898\pm0.0042$ & $1.0066\pm0.0097$ &  $0.9908\pm0.0043$ \\
0.27 & $0.9968\pm0.0069$ & $1.0260\pm0.0021$ & $1.0017\pm0.0078$ &  $1.0246\pm0.0022$ \\
0.28 & $1.0004\pm0.0001$ & $0.9864\pm0.0015$ & $1.0006\pm0.0050$ &  $0.9871\pm0.0016$ \\
\bottomrule
\end{tabular}
\end{center}
\label{tab:Rates}
\end{table}


\subsection{Neutron emissivities and spectra in DRESS} 
\begin{figure}[!bth]
\centering
\includegraphics[scale=0.5]{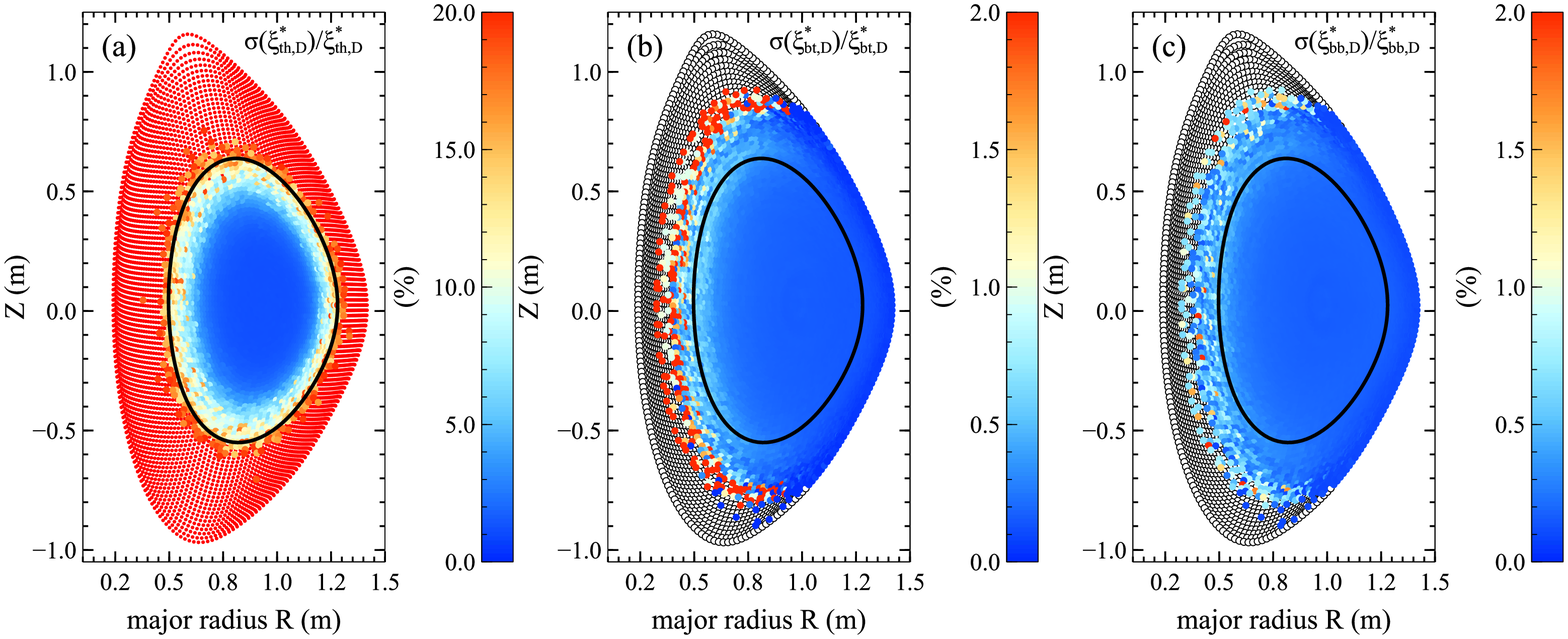} 
\caption{Pulse 29909: fractional errors: a) $\sigma(\xi^*_{\mathrm{th,D}})/\xi^*_{\mathrm{th,D}}$, b) $\sigma(\xi^*_{\mathrm{bt,D}})/\xi^*_{\mathrm{bt,D}}$ and c) $\sigma(\xi^*_{\mathrm{bb,D}})/\xi^*_{\mathrm{bb,D}}$ computed by DRESS at $t = 0.216$ s. The flux surface with $\rho_{\phi} = 0.575$ shown in black encloses the plasma region from which $99\%$ of all neutrons are emitted. The fractional errors exceeding $20\%$ are plotted in red while spatial points from which no neutrons are emitted are shown in white.}\label{fig:DRESS_TH_BT_BB_fractional_errors}
\end{figure}

The DRESS code calculates the thermonuclear, beam-thermal and beam-beam neutron emissivities according to equation~\ref{emissivity} but unlike TRANSP and NUBEAM, it uses the differential fusion reaction cross-section obtained by combining the parameterization from~\cite{BOSCH_HALE} for the total fusion reaction cross section and a Legendre polynomial expansion from the ENDF database~\cite{ENDF} for the angular dependence. DRESS calculations of the neutron emissivities $\xi^*_{x,\mathrm{D}}$, rates $R_{x, \mathrm{D}}$ (where index ``D'' denotes DRESS) and neutron energy spectra presented in this work use as inputs $f_\textrm{T}(\mathrm{R,Z}, E,\lambda)$ and $f_\textrm{A}(\mathrm{R,Z}, E,\lambda)$ (where indexes ``T'' and ``A'' denotes TRANSP and ASCOT), the kinetic profiles $T_{\mathrm{i}}$, $n_{\mathrm{d}}$ and the plasma rotation $\omega$.
The thermal ion velocity distribution is internally modelled in DRESS as a Maxwellian with the local temperature $T_{\mathrm{i}}$. DRESS evaluates the thermonuclear $\xi^*_{\mathrm{th,D}}$, the beam-thermal $\xi^*_{\mathrm{bt,D}}$ and the beam-beam $\xi^*_{\mathrm{bb,D}}$ neutron emissivity components on the same spatial grid on which $f_T$ and $f_A$ are calculated by NUBEAM and ASCOT allowing a point to point comparison between the calculated neutron emissivities. The statistical uncertainties on the estimated neutron emissivities, rates and spectra are also calculated in DRESS. The fractional uncertainties on the $\xi^*_{\mathrm{th,D}}$, $\xi^*_{\mathrm{bt,D}}$ and $\xi^*_{\mathrm{bb,D}}$ by using $f_T$ are presented in figure~\ref{fig:DRESS_TH_BT_BB_fractional_errors}. The fractional uncertainties on the $\xi^*_{\mathrm{th,D}}$ and on the $\xi^*_{\mathrm{bt,D}}$ and $\xi^*_{\mathrm{bb,D}}$ are generally smaller than $2\%$ and $0.5\%$ for $\rho_{\phi} \leq 0.575$, respectively. 
Generally, the fractional uncertainties on the emissivities predicted by DRESS are an order of magnitude smaller than the ones obtained from TRANSP/NUBEAM. This is due to the fact that DRESS is more efficient than NUBEAM for this kind of calculations in terms of computational time.

\begin{figure}[!bth]
\centering
\includegraphics[scale=0.6, trim = 0cm 0cm 0cm 0cm, clip=true]{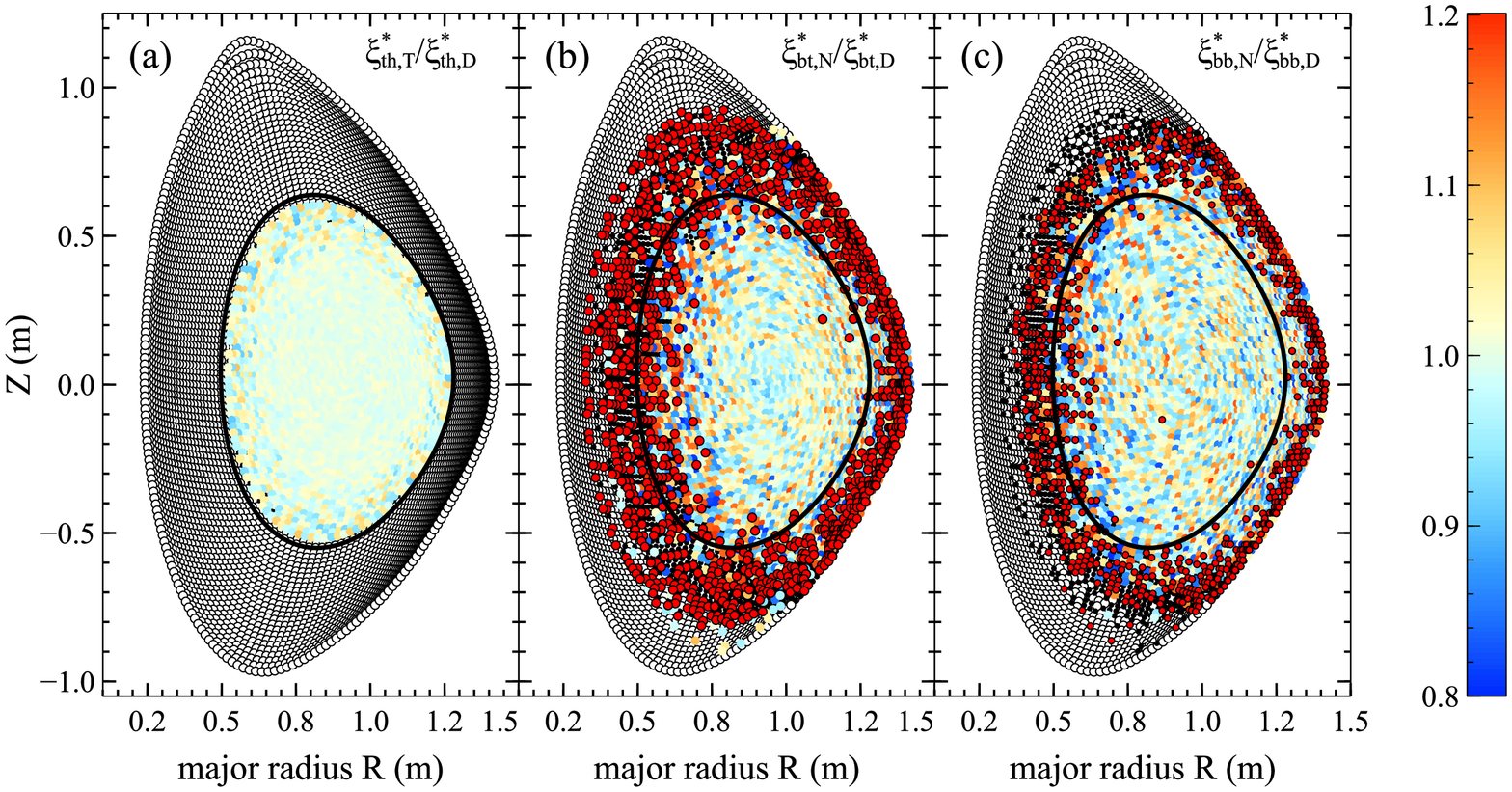}
\caption{Pulse 29909: ratio of TRANSP/NUBEAM to DRESS calculated a) thermonuclear, b) beam-thermal and c) beam-beam emissivities at $t = 0.216$ s. The flux surface restricting the plasma region from which $99\%$ of all neutrons are emitted is shown in black. The ratios exceeding $20\%$ are plotted in red while spatial points for which the evaluated ratio is zero or below $80\%$ are plotted in white.}
\label{fig:DRESS_TH_BT_BB_emissivity}
\end{figure}

\begin{figure}[!bth]
\centering
\includegraphics[scale=0.57, trim = 0cm 0cm 0cm 0cm, clip=true]{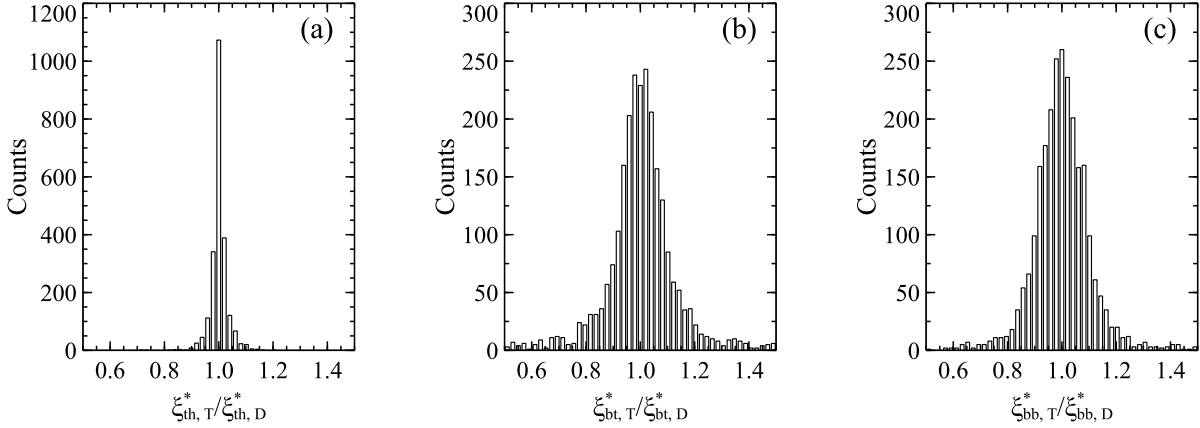} 
\caption{Pulse 29909: Histograms formed from the distributions of ratios of (a) $\xi^*_{\mathrm{th,T}}/ \xi^*_{\mathrm{th,D}}$, (b) $\xi^*_{\mathrm{bt,T}}/ \xi^*_{\mathrm{bt,D}}$ and (c) $\xi^*_{\mathrm{bb,T}}/ \xi^*_{\mathrm{bb,D}}$ for $\rho_{\phi} \leq 0.575$ at $t = 0.216$ s.}\label{fig:2D_TH_BT_BB_emiss_R_TD_HISTOGRAM}
\end{figure}

\begin{figure}[!bth]
\centering
\includegraphics[scale=0.57, trim = 0cm 0cm 0cm 0cm, clip=true]{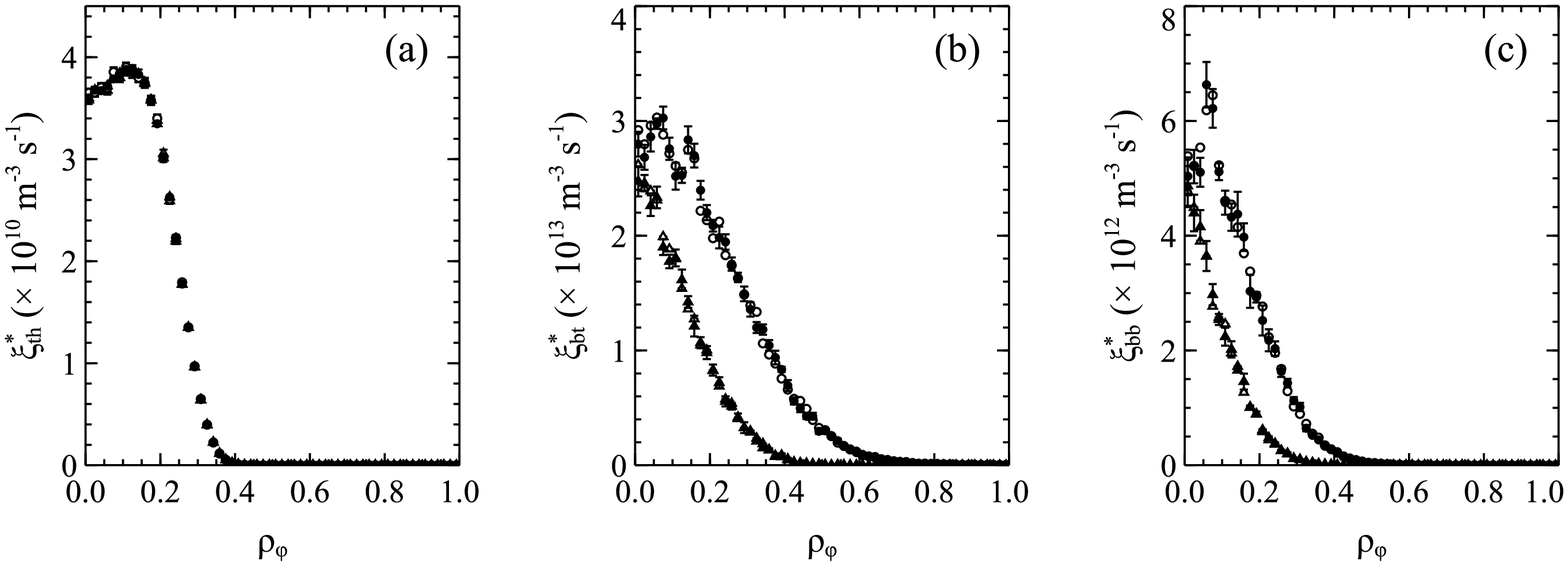} 
\caption{Pulse 29909: non-flux averaged (a) thermonuclear, (b) beam-thermal and (c) beam-beam neutron emissivities along the plasma mid-plane predicted by TRANSP/NUBEAM (full circles and triangles) and DRESS (open circles and triangles) at $t = 0.216$ s. The circles and triangles represent the neutron emissivities calculated along the mid-plane for points going from the plasma center to outboard and inboard side, respectively.}\label{fig:DRESS_TH_BT_BB_emissivity_absolute_scale}
\end{figure}

\subsection{Benchmark of DRESS against TRANSP/NUBEAM}
\label{bench}

\begin{figure}
\centering
\includegraphics[scale=0.70, trim = 0cm 0cm 0cm 0cm, clip=true]{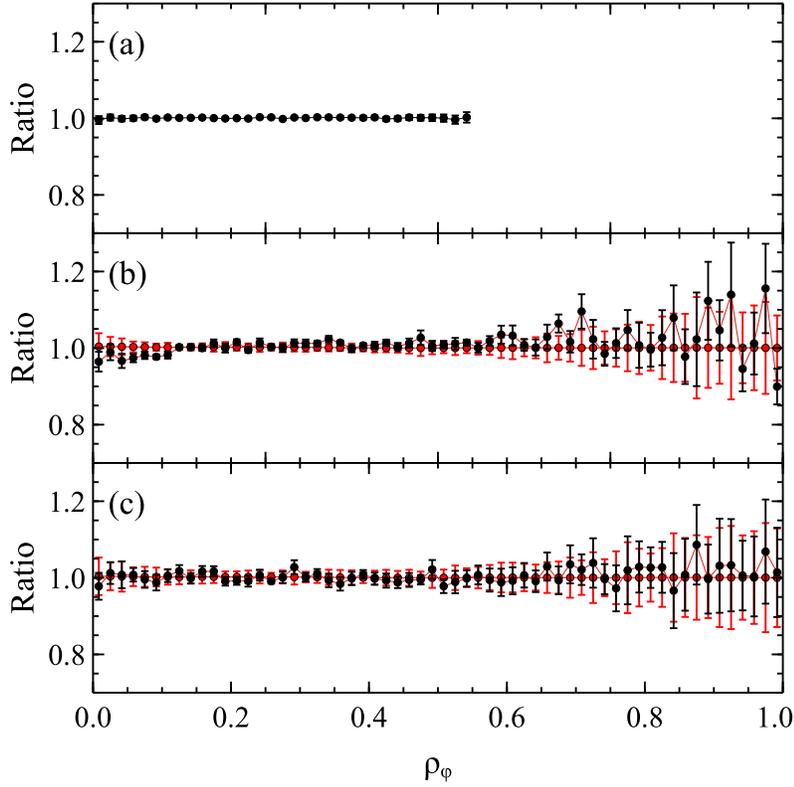}

\caption{Pulse 29909 : ratio of (a) $\xi_{\mathrm{th,D}}$ to $\xi_{\mathrm{th,T}}$, (b) $\xi_{\mathrm{bt}}$ to $\xi_{\mathrm{bt,T}}$ and (c) $\xi_{\mathrm{bb}}$ to $\xi_{\mathrm{bb,T}}$ for the time slice $t = 0.216$ s as a function $\rho_{\phi}$. The ratio between $\xi_{x}$ and $\xi_{x,\mathrm{T}}$ based on $\xi^*_{x,\mathrm{N}}$ are shown in red while the ratio between $\xi_{x,\mathrm{D}}$ and $\xi_{x,\mathrm{T}}$ based on $\xi^*_{x,\mathrm{D}}$ are shown in black (index ``$x$'' denotes the thermonuclear, beam-thermal or beam-beam emissivity component while index ``D'' stands for DRESS). TRANSP sets $\xi_{\mathrm{th,T}}$ to zero as a result of $T_{\mathrm{i}}$ being lower than 0.2 keV for $\rho_{\phi} \geq 0.575$, hence the calculated ratio of $\xi_{\mathrm{th,T}}$ to $\xi_{\mathrm{th,D}}$ is also equal to zero.}\label{fig:Ratios}
\end{figure}

The benchmark of DRESS against TRANSP/NUBEAM has been done in three steps. Firstly, the ratio of the non-flux averaged thermonuclear, beam-thermal and beam-beam emissivities calculated by TRANSP/NUBEAM to DRESS has been evaluated and is shown in figure~\ref{fig:DRESS_TH_BT_BB_emissivity}. The estimations of the thermonuclear, beam-thermal and beam- beam emissivities by TRANSP/NUBEAM and DRESS agree within $20\%$ in the region $\rho_{\phi} \leq 0.575$. The ratios of TRANSP/NUBEAM to DRESS predictions larger than $120\%$ (shown as red data points) and smaller than $80\%$ (shown as white data point), which arose due to large statistical fluctuations of $\xi^*_{\mathrm{x,T}}$, are mainly observed for $\rho_{\phi} \geq 0.575$. However, the contribution from this region to the neutron emissivity components is negligible (less than $1\%$) and therefore does not significantly contribute to the total neutron emission, even taking into account the large uncertainties. Figures \ref{fig:2D_TH_BT_BB_emiss_R_TD_HISTOGRAM} and \ref{fig:DRESS_TH_BT_BB_emissivity_absolute_scale} shown the histograms of the ratios between the neutron emissivity components calculated by TRANSP/NUBEAM and DRESS, and the non-flux averaged neutron emissivity profiles as a function of $\rho_{\phi}$ calculated by TRANSP/NUBEAM (full circles and triangles) and DRESS (open circles and triangles). As can be seen, the TRANSP/NUBEAM and DRESS estimations of emissivities are indeed consistent with each other within the statistical uncertainties.\\
Secondly, a component by component comparison between the flux averaged $\xi_{x\mathrm{,T}}$ and $\xi_{x\mathrm{,D}}$ has also been carried out. The $\xi_{x\mathrm{,D}}$ has been obtained by averaging the corresponding $\xi^*_{x\mathrm{,D}}$ over poloidal zones for each flux surface using the following expression 
\begin{equation}
\xi_{x}({\rho_{\phi}}_{i})= \frac{\sum_{j = 2i^2 - 2i + 1}^{2(i^2 + i)} \xi^*_{x}(\mathrm{R}_{i,j},\mathrm{Z}_{i,j};{\rho_{\phi}}_{i}) \mathrm{d}V_{i,j}(\mathrm{R}_{i,j},\mathrm{Z}_{i,j};{\rho_{\phi}}_{i})}{\sum_{j = 2i^2 - 2i + 1}^{2(i^2 + i)} \mathrm{d}V_{i,j}(\mathrm{R}_{i,j},\mathrm{Z}_{i,j};{\rho_{\phi}}_{i})},\label{eq:emissivity}
\end{equation}
where $i$ and $j$ are the annular and the poloidal zone indexes and $\mathrm{d}V_{i,j}$ is the corresponding volume. This equation has been verified by comparing $\xi_{x, \mathrm{T}}$ with $\xi_{x}$ computed using the non-flux averaged beam-thermal and beam-beam emissivity components provided by NUBEAM for one selected time slice. The ratio between $\xi_{x}$ and $\xi_{x,\mathrm{T}}$ for the beam-thermal and beam-beam component is very close to one, as can be seen in panels b) and c) of figure~\ref{fig:Ratios}, indicating that equation~\ref{eq:emissivity} is correct and can be used to calculate flux-averaged neutron emissivity. 
The $\xi_{x\mathrm{,D}}$ computed from $\xi^*_{x\mathrm{,D}}$ are also consistent with the $\xi_{x\mathrm{,T}}$ within the statistical uncertainties.   

Finally, the thermonuclear $R_{\mathrm{th,D}}$, beam-thermal $R_{\mathrm{bt,D}}$ and beam-beam $R_{\mathrm{bb,D}}$ neutron rates for few selected times were obtained by integrating the $\xi^*_{\mathrm{th,D}}$, $\xi^*_{\mathrm{bt,D}}$ and $\xi^*_{\mathrm{bb,D}}$ over the plasma volume to allow a comparison with TRANSP estimations.  
The $R_{\mathrm{th,D}}$, $R_{\mathrm{bt,D}}$, $R_{\mathrm{bb,D}}$ along with total rate $R_{D}$ evaluated for several time slices are summarized in table~\ref{tab:Rates} and shown in figure~\ref{fig:Rates}. The ratio of TRANSP to DRESS computed rates together with estimated uncertainties are also presented in Table~\ref{tab:Rates}. Generally, the agreement between DRESS and TRANSP predicted rates is very good as the calculated ratios are equal to one within statistical uncertainties for all selected time slices. This concludes the benchmarking of DRESS against TRANSP/NUBEAM. In the next section, DRESS will be used to model the neutron spectra at the detector location thus allowing to predict the expected count rates at the detectors.

\section{Modelling of the deposited neutron energy in the EJ301 liquid scintillation detectors}
\label{sec:NeutronSpectra}
On MAST time-resolved neutron emission profiles and energy deposited spectra are measured by the NC which is equipped with four LoS, two lying on the equatorial plane and two inclined which look 20 cm below the equatorial plane. Each LoS is equipped with liquid scintillator of the type EJ301. The light output from liquid scintillators exposed to a beam of mono-energetic neutrons is dominated by the de-excitation of the detector molecules excited by recoil protons resulting from the elastic collision of the neutrons with the Hydrogen atoms. All scattering angles are roughly equiprobable and the recoil proton pulse height spectrum (PHS) extends from zero energy (grazing collision) to the incident neutron energy (head-on collision). The ideal box-like recoil proton PHS is smoothed at the high energy end and enhanced at the low energy end by the detector finite energy resolution and by its non-linear light output \cite{CECCONRESP}. In addition, since neutrons emitted from a plasma are not mono-energetic, modelling of the experimentally measured PHS requires the estimation of the neutron energy spectrum $\Psi$ at the detector location and of the response function matrix of the liquid scintillator for all neutron energies of interest.

\subsection{Neutron spectrum}\label{sec:NS}
The energy of a neutron created in a fusion reaction depends on the velocities of the interacting fuel ions. If interacting ions are in thermal equilibrium in a stationary plasma, the resulting neutron energy spectrum is approximately a Gaussian centered on the mean neutron energy $E_{\mathrm{n}} = 2.45$ MeV and with a full width at half maximum proportional to square root of ion temperature $(\mathrm{FWHM} \propto \sqrt{T_\mathrm{i}})$. On the other hand, if the plasma is rotating or if at least one of ions participating in a fusion reaction has an anisotropic velocity distribution, the shape of the generated neutron energy spectrum depends strongly on the observation direction. A neutron spectrum observed in a direction perpendicular to the magnetic field has the typical ``double humped'' shape, as the neutron energy will be positively or negatively Doppler shifted when the fast ion moves towards or away from the detector during a gyro-orbit. Another typical situation is when the neutron source is observed in a direction parallel to the magnetic field (tangential LoS): in this case, due to the Doppler effect, the neutron energy spectrum will be shifted to higher or lower energies depending on whether the fast ion moves towards or away from the detector. 

\begin{figure}[!bth]
\centering
\includegraphics[scale=0.5]{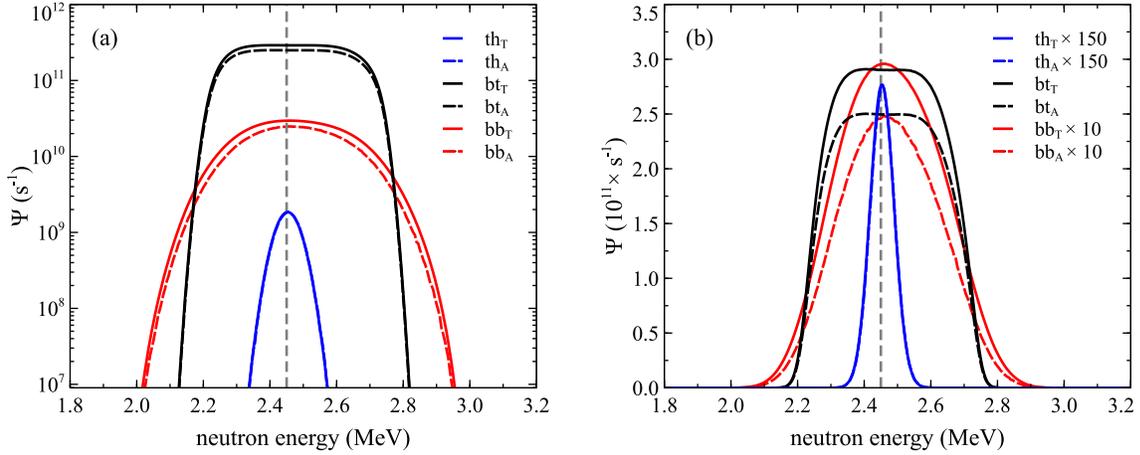}  
\caption{Pulse 29909: thermonuclear (blue), beam-thermal (black) and beam-beam (red) un-collided neutron energy spectra calculated by DRESS and integrated over the entire plasma volume at time $t = 0.216$ s shown in a) logarithmic and b) linear scale for TRANSP (solid lines) and ASCOT (dashed lines) inputs. The vertical dashed grey line indicates $E_{\mathrm{n}} = 2.45$ MeV.}\label{fig:NS_direct_entire_plasma.png}
\end{figure}

\begin{figure}[!bth]
	\centering
	\includegraphics[scale=0.50, trim = 0cm 0cm 0cm 0cm, clip=true]{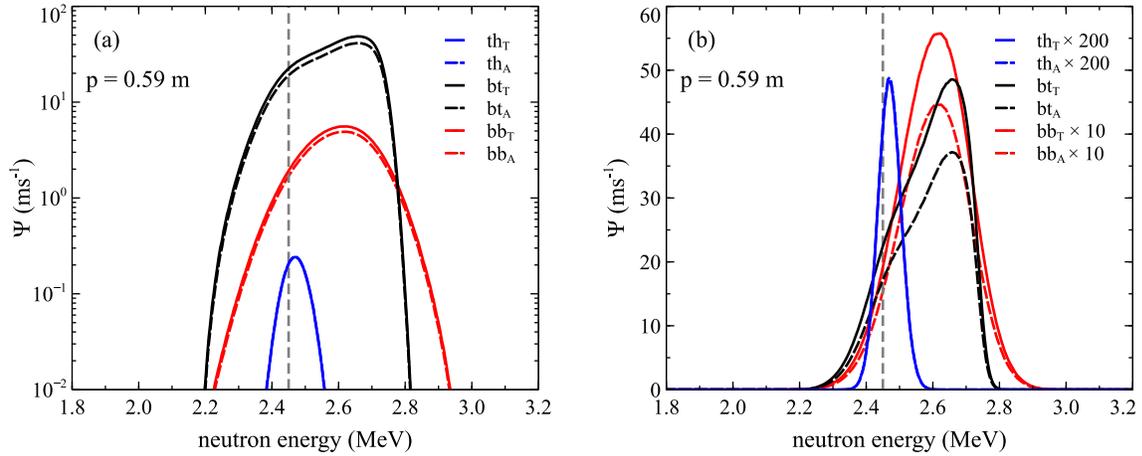}
	\caption{Pulse 29909: thermonuclear (blue), beam-thermal (black) and beam-beam (red) neutron energy spectra calculated by DRESS and integrated within the field of view of a detector set at tangency radius $p = 0.59$ m at time $t = 0.216$ s shown in a) logarithmic and b) linear scale for TRANSP (solid lines) and ASCOT (dashed lines) inputs. The vertical dashed grey line indicates $E_{\mathrm{n}} = 2.45$ MeV.}\label{fig:NS_singleLOS.png}
\end{figure}

An example of the un-collided thermonuclear $\Psi_{\mathrm{th}}$, beam-thermal $\Psi_{\mathrm{bt}}$ and beam-beam $\Psi_{\mathrm{bb}}$ neutron energy spectra integrated over the entire $4\pi$ solid angle is shown in figure~\ref{fig:NS_direct_entire_plasma.png} for MAST pulse 29909 at $t=0.216$ s. In order to study the effect of GC approximation neutron energy spectra have been calculated by DRESS using the fast ion distributions calculated by TRANSP/NUBEAM and ASCOT/BBNBI. The $\Psi_{\mathrm{th}}$ component peaks at $E_{\mathrm{n}} \simeq 2.46\ \mathrm{MeV}$ and this slight upwards shift respect to $E_n = 2.45$ MeV is due to the isotropy in the centre of mass system in equation (29) of \cite{BRYSK} which results in a shift of about 10 keV \cite{JAN}. The $\Psi_{\mathrm{bt}}$ component dominates in the energy region $2.1 \leq E_{\mathrm{n}} \leq 2.8$ MeV and peaks at $E_{\mathrm{n}} \simeq 2.41$ MeV. Outside this region, the $\Psi_{\mathrm{bb}}$ (peaking at $E_{\mathrm{n}} \simeq 2.47\ \mathrm{MeV}$) overcomes $\Psi_{\mathrm{bt}}$ due to the broader range of the interacting ion relative velocities. The $\Psi_{\mathrm{th}}$ and $\Psi_{\mathrm{bb}}$ components are symmetric with respect to mean neutron energy, whereas the $\Psi_{\mathrm{bt}}$ component is slightly asymmetric due to the plasma rotation, with spectra from TRANSP/NUBEAM and ASCOT fast ion distributions having the same shape. However, the amplitude of the spectra calculated by ASCOT/BBNBI is about 20 \% lower than the one computed by TRANSP/NUBEAM due to shorter slowing down times for the fast ions in GO mode.\\
The neutron energy spectra shown in figure~\ref{fig:NS_direct_entire_plasma.png} can be compared with the neutron energy spectra calculated in the direction of a specific NC detector. An example of the collimated thermonuclear $\Psi_{\mathrm{th}}\left( p \right) $, beam-thermal $\Psi_{\mathrm{bt}}(p)$ and beam-beam $\Psi_{\mathrm{bb}}(p)$ neutron energy spectra incident on a detector looking at a tangency radius $p = 0.59$ m is shown in figure~\ref{fig:NS_singleLOS.png}.\\
At this location, the majority of the fast ions moves towards the detector and a strong upward Doppler shift of neutron energy distribution, expected from theoretical considerations, is indeed observed. As can be seen, $\Psi_{\mathrm{bt}}$  peaks at $E_{\mathrm{n}} \simeq 2.65\ \mathrm{MeV}$ reflecting the fact that neutron emission is dependent on the plasma rotation direction. For $E_\mathrm{n} \geqslant 2.8$ MeV, $\Psi_{\mathrm{bb}}$ dominates over $\Psi_{\mathrm{bt}}$ due to the wider velocity range accessible for interacting fast ions. The $\Psi_{\mathrm{th}}$ is symmetric and its maximum is also Doppler shifted to energy $E_{\mathrm{n}} \simeq 2.47\ \mathrm{MeV}$ as a results of plasma rotation towards the detector. Examples of the collimated neutron energy spectrum components for different tangency radii are presented in figures~\ref{fig:NS_singleLOS.png} and \ref{fig:NS_HD_different_LoS.png}. 

\subsection{Measured and modelled proton pulse height spectra}\label{sec:PHS}

\begin{figure}[!bth]
\centering
\includegraphics[scale=0.6]{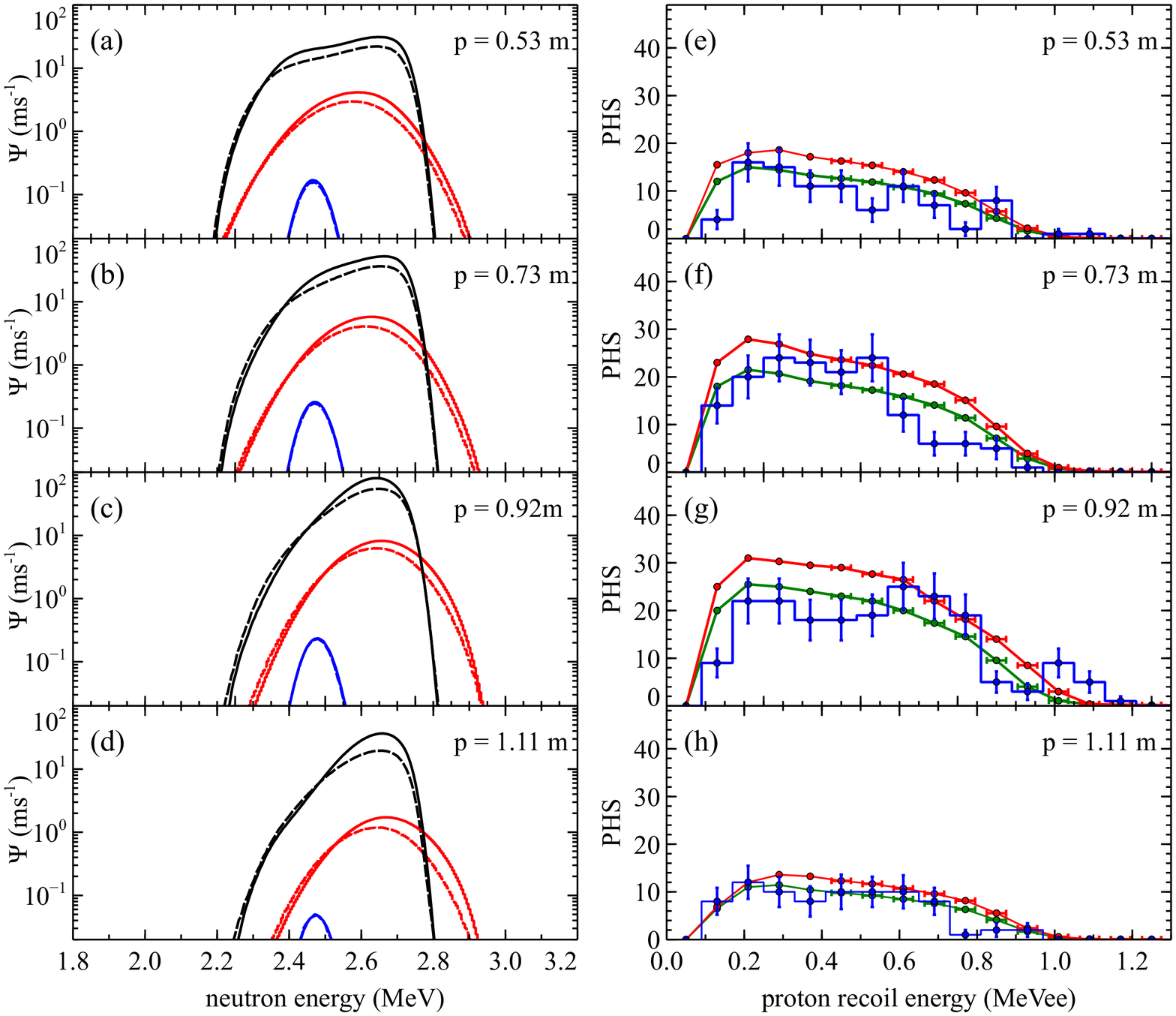} 
\caption{Collimated thermonuclear (blue), beam-thermal (black) and beam-beam (red) neutron energy spectra for TRANSP/NUBEAM - DRESS (solid lines) and ASCOT/BBNBI - DRESS (dashed lines) are plotted in panels a) - d) for different tangency radii $p$. The modelled from TRANSP/NUBEAM - DRESS (red points with the solid red line used to guide eyes), from ASCOT/BBNBI - DRESS (green points with the solid geeen line used to guide eyes) and measured (blue points) proton recoil pulse height spectra for corresponding neutron energy spectra shown in panels a) - d) are presented in panels in e) - h). The vertical blue error bars represent the statistical uncertainty associated with the measured PHS while the horizontal red and green error bars represent the uncertainty associated with the light output function used to estimate the modelled PHS.}\label{fig:NS_HD_different_LoS.png}
\end{figure}
The neutron energy spectrum is converted into a recoil proton energy spectrum via n-H elastic scattering. This is then converted into a light pulse height spectra which is measured. This conversion is referred to as the detector response function. The PHS is further affected by the threshold of acquisition system, ``pile-up'' events and scattered neutrons. The response function matrix of the EJ301 detector to mono-energetic neutrons has been calculated with the NRESP code~\cite{NRESP} in the range from 1 to 3.5 MeV in steps of 5 keV. The response function of the cylindrical approximation of the NC detector in NRESP has been benchmarked against a realistic NC detector geometry for selected neutron energies using MCNP \cite{MCNP}. The NRESP code requires as inputs the detector energy resolution and the proton light output function. The detector energy resolution has been experimentally determined for one out of the four EJ301 detectors and it is here assumed to be the same for all detectors used in this work~\cite{Cecconello}. The proton light output function found in literature~\cite{Verbinski} has been used in this work.\\
The acquisition threshold has been included in the modelling of the PHS after folding the total neutron spectrum with the response function matrix based on the acquisition of $^{22}$Na $\gamma-$ray spectra and estimated to be $E_{\mathrm{thr}} \simeq 0.12$ MeVee $(1 \mathrm{MeVee} \simeq 2.86\ \mathrm{MeV\ proton})$. The contributions from ``pile-up'' (containing two or more superimposed separable events) and coincident ``pile-up'' events in the detectors have been investigated. The former are accounted for in the NC PHS by means of the analysis code described in \cite{Cecconello}. The coincident ``pile-up'' events are, on the other hand, events that cannot be distinguished from single events because they are generated in the detector on a time scale shorter than the fastest sampling rate of the acquisition system. Although these events are present in the entire PHS spectrum, they are best seen in the high energy tail of the PHS. Such coincident ``pile-up'' events are clearly seen in the experimental PHS measured at $p = 0.92$ m above the energy 1 MeVee. The contribution of the coincident ``pile-up'' events to the modelled PHS, based on theoretical considerations~\cite{KNOLL}, has been estimated to be $0.8\%$, $1.1\%$, $1.4\%$ and $0.7\%$ for LoS shown in panels $\mathrm{(e)-(h)}$ of figure~\ref{fig:NS_HD_different_LoS.png}, leading to the conclusion that they do not make a significant contribution to the measured PHS.\\
In this study the scattered neutron contribution to the energy spectra, unavoidably present in the experimental data, are not included in the modelling. This contribution is typically small in the plasma central region $ p \simeq 0.9$ m, nevertheless it becomes important for tangency radii $p\leq 0.25$ m and $p \geq 1.15$ m \cite{Cecconello}. The PHS here shown were measured for a plasma region $0.53\leq p\leq 1.11$ m where the ratio between scattered and direct neutron is about $\simeq 4$ \% and therefore the scattered neutrons do not contribute significantly to the PHS. The comparison of the modelled and measured PHS for different LoS is depicted in panels $\mathrm{e)-h)}$ of figure~\ref{fig:NS_HD_different_LoS.png}. The modelled PHS have been fitted to the experimental ones using Cash statistic ($C$)~\cite{Cstat} where the amplitude of the PHS is proportional to the scaling factor $k$ and is the only one free fitting parameter. The best fits to the experimental PHS and their reduced $C_{\mathrm{red}}$ for TRANSP/NUBEAM and ASCOT/BBNBI fast ion distributions are reported in table \ref{final}. Clearly, a better agreement between the measured PHS and the synthetic ones calculated using ASCOT fast ion distribution than the one from TRANSP/NUBEAM has been obtained. This is due to the fact that the GO is more suitable for this kind of calculations on MAST than the GC approximation in TRANSP/NUBEAM, since it is more suitable for the calculation of the fast ion transport simulations in spherical tokamaks due to the combination of a low $\textbf{B}$ field and a large gradient $\nabla \textbf{B}$.

\begin{table}[!bth]
	\begin{center}
			\caption{Estimate of the fit parameters $k$ and their $C_{\mathrm{red}}$ for each impact parameter $p$.}
			\renewcommand{\arraystretch}{1.15}
	\begin{tabular}{ccc|cc}
		\hline
		\multicolumn{1}{c}{} &
		\multicolumn{2}{c|}{TRANSP/NUBEAM} &
		\multicolumn{2}{c}{ASCOT/BBNBI} \\
		$p$ (m)& $k$ & $C_{\mathrm{red}}$ & $k$ & $C_{\mathrm{red}}$  \\
		\hline
		0.53 & 0.69 & 1.89 & 0.89 & 1.86  \\
		\hline
		0.73 & 0.72 & 1.74 & 0.94 & 1.68  \\
		\hline
		0.92 & 0.71 & 4.67 & 0.85 & 3.95  \\
		\hline
		1.11 & 0.76 & 1.14 & 0.92 & 0.95  \\

		\hline
	\end{tabular}
	\label{final}
		\end{center}
\end{table}

\section{Summary and Conclusions}
\label{conclusions}
A series of TRANSP/NUBEAM simulations have been performed for a MHD-quiescent MAST plasma discharge to estimate the statistical fluctuations of the predicted neutron emissivities and rates. These TRANSP/NUBEAM simulations provided the necessary input data for the DRESS code and allowed for thorough validations of neutron emission calculation in DRESS against numerical calculations and measurements. This validation has been performed in two steps. Firstly, the neutron emissivities and rates computed by DRESS were compared with the ones predicted by TRANSP/NUBEAM codes. Excellent agreement between DRESS and TRANP/NUBEAM estimation of emissivities and rates was found. Secondly, the same TRANSP/NUBEAM simulations and ASCOT/BBNBI in Gyro-Orbit ones were used by DRESS to model the collimated neutron energy spectrum at different position of detectors employed by the NC. The collimated and scattered neutron energy spectra were summed, convoluted with the detector's response function matrix and then compared with the recoil proton PHS measured by the NC. Good agreement is found between the simulated PHS from TRANSP/NUBEAM when taking into account the same scaling factor used in~\cite{MASTDEFICIT}. Instead, a better agreement is obtained between the ASCOT/BBNBI in Gyro-Orbit PHS and the measured ones, requiring a scaling factor around 0.9. An accompanying paper will discuss possible reasons for the need of this scaling factor.

\ack
This work was funded by the Swedish Research Council, the RCUK Energy Programme under grant EP/I501045, authored in part by Princeton University under Contract No. DE-AC02-09CH11466 and is based upon the work supported by the U.S. Department of Energy, Office of Science, Office of Fusion Energy Sciences, and has been carried out within the framework of the EUROfusion Consortium and has received funding from the Euratom research and training programme 2014-2018 and 2019-2020 under grant agreement No 633053. The views and opinions expressed herein do not necessarily reflect those of the European Commission.

\section*{References}
\bibliographystyle{iopart-num} 
\bibliography{Bibliografia}

\end{document}